\documentclass[12pt,a4paper]{article}

\usepackage{epsfig}
\usepackage{amsmath,amsfonts,amssymb}
\usepackage{cite}
\usepackage{colortbl}
\usepackage{pifont}

\providecommand{\openone}{\leavevmode\hbox{\small1\kern-3.8pt\normalsize1}}

\newcommand{\gm}{\gamma^\mu}
\newcommand{\smn}{\sigma^{\mu \nu}}
\newcommand{\Wmn}{W_{\mu \nu}}
\newcommand{\Bmn}{B_{\mu \nu}}

\newcommand{\dvz}{d_V^Z}
\newcommand{\daz}{d_A^Z}
\newcommand{\dva}{d_V^\gamma}
\newcommand{\daa}{d_A^\gamma}
\newcommand{\vl}{V_L}
\newcommand{\vr}{V_R}
\newcommand{\gl}{g_L}
\newcommand{\gr}{g_R}

\newcommand{\vij}{\mathcal{V}_{ij}}
\newcommand{\tij}{\mathcal{T}_{ij}}
\newcommand{\vLL}{\mathcal{V}_{LL}}    
\newcommand{\vLR}{\mathcal{V}_{LR}}  
\newcommand{\vRL}{\mathcal{V}_{RL}}  
\newcommand{\vRR}{\mathcal{V}_{RR}}  
\newcommand{\tLV}{\mathcal{T}_{LV}}    
\newcommand{\tLA}{\mathcal{T}_{LA}}    
\newcommand{\tRV}{\mathcal{T}_{RV}}    
\newcommand{\tRA}{\mathcal{T}_{RA}}    

\newcommand{\RE}{\text{Re}}
\newcommand{\IM}{\text{Im}}

\begin{document}

\begin{center}
\begin{Large}
{\bf Top effective operators at the ILC}
\end{Large}

\vspace{0.5cm}
J. A. Aguilar--Saavedra$^{a,b}$, M. C. N. Fiolhais$^c$, A. Onofre$^d$ \\[1mm]
\begin{small}
{\it $^a$ Departamento de F\'{\i}sica Te\'orica y del Cosmos, 
Universidad de Granada, Granada, Spain} \\
{\it $^b$ Instituto de F\'{\i}sica de Cantabria (CSIC-UC), Santander, Spain} \\
{\it $^c$ LIP, Departamento de Fisica, Universidade de Coimbra,  Coimbra, Portugal} \\
{\it $^d$ LIP, Departamento de Fisica, Universidade do Minho, Braga, Portugal}
\end{small}
\end{center}

\begin{abstract}
We investigate the effect of top trilinear operators in $t \bar t$ production at the ILC. We find that the sensitivity to these operators largely surpasses the one achievable by the LHC either in neutral or charged current processes, allowing to probe new physics scales up to 4.5 TeV for a centre of mass energy of 500 GeV. We show how the use of beam polarisation and an eventual energy upgrade to 1 TeV allow to disentangle all effective operator contributions to the $Ztt$ and $\gamma tt$ vertices.
\end{abstract}

\section{Introduction}

Precision measurements are an essential complement of direct searches for new physics beyond the Standard Model (SM). The most remarkable successes of this interplay may be the prediction of the existence of the charm quark due to the absence of flavour-changing neutral currents, and the prediction of the top quark mass before its actual discovery. With this philosophy, the construction of a high-energy $e^+ e^-$  International Linear Collider (ILC) has been proposed to complement direct searches carried out at the Large Hadron Collider (LHC). In the case of the top quark, precision measurements of its properties, in particular of its couplings, are specially interesting because it is the heaviest elementary particle yet discovered, and as such it is expected to be more sensitive to new physics at higher scales. 

In its first two years of operation, the LHC has not shown any sign of new physics yet. This implies that new particles coupling to the SM ones have masses above the electroweak symmetry breaking scale $v=246$ GeV, and perhaps beyond the TeV. Therefore, the effect of these new degrees of freedom in the top quark properties can be parameterised in terms of dimension-six gauge-invariant effective operators~\cite{Buchmuller:1985jz,AguilarSaavedra:2008zc}.
The difference between this framework and previous approaches for the study of $e^+ e^- \to t \bar t$ at the ILC with anomalous top couplings~\cite{Grzadkowski:1996kn,Abe:2001nqa,Devetak:2010na} is that here we make use of the full $\text{SU}(3)_c \times \text{SU}(2)_L \times \text{U}(1)_Y$ gauge symmetry of the SM, not only the unbroken $\text{SU}(3)_c \times \text{U}(1)_\text{em}$. This larger symmetry leads to several interesting implications. First of all, the effective operator framework allows to reduce the number of independent parameters entering fermion trilinear interactions to four at most~\cite{AguilarSaavedra:2008zc}, one half of the total number of parameters involved in a general off-shell form factor. Second, it allows to set relations between new physics contributions to the top quark interactions, for example between left-handed contributions to the the $Wtb$ and $Ztt$ vertices. Such relations not only reduce further the number of arbitrary parameters, but also introduce an useful synergy between measurements of different top quark vertices. Last, but not least,  this framework is also very convenient since it allows to consistently compute radiative corrections and study the effect of anomalous top interactions in loop observables~\cite{Zhang:2010dr}.

In this paper we study the effect of top trilinear effective operators in $e^+ e^- \to t \bar t$ at the ILC. Our estimates will show that the ILC sensitivity will largely surpass the one achievable at the LHC, either in top decays (current one~\cite{Aad:2012ky} or envisaged~\cite{AguilarSaavedra:2007rs}) or in $t \bar t Z$ and $t \bar t \gamma$ production~\cite{Baur:2004uw}.
Moreover, our focus is not only in the sensitivity to these operators, but rather on discussing how the different ILC beam polarisation options and CM energies could allow to disentangle the various effective operator contributions to the $Ztt$ and $\gamma tt$ vertices. Effective operators can also affect the top decay. However, since the study and the observables involved are the same as for the LHC~\cite{Kane:1991bg,AguilarSaavedra:2006fy,AguilarSaavedra:2010nx}, we defer their study to future work.

The structure of this paper is as follows. In section~\ref{sec:2} we write down the dimension-six operators involved in the $Ztt$ and $\gamma tt$ interactions, which contribute to $e^+ e^- \to t \bar t$ and obtain expressions for the polarised cross sections and asymmetries. In section~\ref{sec:3} we compare the ILC sensitivity to single operators with present or expected LHC limits. The possibility to disentangle these operator contributions is discussed in detail in section~\ref{sec:4}. In section~\ref{sec:5} we draw our conclusions.

\section{$e^+ e^- \to t \bar t$ with effective operators}
\label{sec:2}

A minimal non-redundant set of dimension-six operators contributing to top quark vertices was presented in Refs.~\cite{AguilarSaavedra:2008zc}. The operators contributing to $Ztt$ and $\gamma tt$ interactions are only five,
\begin{align}
& O_{\phi q}^{(3,3+3)} = i \left[ \phi^\dagger (\tau^I D_\mu - \overleftarrow
 D_\mu \tau^I) \phi \right] \;
       (\bar q_{L3} \gm \tau^I q_{L3}) \,,
  && O_{uW}^{33} = (\bar q_{L3} \smn \tau^I t_{R}) \tilde \phi \, \Wmn^I \,, \notag \\
& O_{\phi q}^{(1,3+3)} = i (\phi^\dagger \overleftrightarrow D_\mu \phi) (\bar q_{L3} \gm q_{L3}) \,,
  && O_{uB\phi}^{33} = (\bar q_{L3} \smn t_{R}) \tilde \phi \, \Bmn \,, \notag \\
& O_{\phi u}^{3+3} = i (\phi^\dagger \overleftrightarrow D_\mu \phi) (\bar t_{R} \gm t_{R}) \,,
\label{ec:Oall}
\end{align}
using standard notation with $\tau^I$ the Pauli matrices, $q_{L3}$ the left-handed third generation quark doublet, $t_R$ the right-handed top quark singlet, $\phi$ the SM Higgs doublet, $\tilde \phi = i \tau^2 \phi^*$, $\Wmn^I$ and $\Bmn$ the $\text{SU}(2)_L$ and $\text{U}(1)_Y$ field strength tensors, respectively, $D_\mu$ ($\overleftarrow D_\mu$) the covariant derivative acting on the right (left) and $\overleftrightarrow D_\mu = D_\mu - \overleftarrow D_\mu$. The three operators in the left column of Eqs.~(\ref{ec:Oall}) are Hermitian, hence their coefficients are real. Including the SM and dimension-six operator contributions, the $Ztt$ vertex reads
\begin{eqnarray}
\mathcal{L}_{Ztt} & = & - \frac{g}{2 c_W} \bar t \, \gm \left( c_L^t P_L
  + c_R^t P_R \right) t\; Z_\mu  - \frac{g}{2 c_W} \bar t \, \frac{i \smn q_\nu}{M_Z}
  \left( \dvz + i \daz \gamma_5 \right) t\; Z_\mu  \,,
\label{ec:Ztt}
\end{eqnarray}
with $c_L^t = X_{tt}^L - 2 s_W^2 Q_t$, $c_R^t = X_{tt}^R - 2 s_W^2 Q_t$ ($Q_t=2/3$ is the top quark electric charge) and
\begin{align}
& X_{tt}^L = 1  + \left[  C_{\phi q}^{(3,3+3)} - C_{\phi q}^{(1,3+3)}
\right]  \frac{v^2}{\Lambda^2} \,,
&& \dvz = {\sqrt 2} \, \RE \left[ c_W  C_{uW}^{33} - s_W C_{uB\phi}^{33} \right]  \frac{v^2}{\Lambda^2}
\,, \notag \\
& X_{tt}^R =  - C_{\phi u}^{3+3} \frac{v^2}{\Lambda^2} \,,
&& \daz = {\sqrt 2} \, \IM \left[ c_W C_{uW}^{33} - s_W  C_{uB\phi}^{33} \right] \frac{v^2}{\Lambda^2}
\,,
\label{ec:Ztt2}
\end{align}
where the $C$ constants are the coefficients of the operators in Eqs.~(\ref{ec:Oall}) and $\Lambda$ is the new physics scale. The $\gamma tt$ vertex reads
\begin{equation}
\mathcal{L}_{\gamma tt} = - e Q_t \bar t \, \gm  t\; A_\mu  - e \bar t \, \frac{i \smn q_\nu}{m_t} \left( \dva + i \daa \gamma_5 \right) t\; A_\mu  \,.
\label{ec:gatt}
\end{equation}
with
\begin{eqnarray}
\dva & = & \frac{\sqrt 2}{e} \, \RE \left[s_W  C_{uW}^{33} + c_W C_{uB\phi}^{33} \right] \frac{v m_t}{\Lambda^2}
  \,, \notag \\
\daa & = & \frac{\sqrt 2}{e} \, \IM \left[s_W  C_{uW}^{33} +  c_W C_{uB\phi}^{33} \right] \frac{v m_t}{\Lambda^2} \,.
\label{ec:gatt2}
\end{eqnarray}
Thus,  the total number of real parameters necessary to describe non-SM contributions to the $Ztt$ and $\gamma tt$ vertices is six, corresponding to five dimension-six operators, three of them Hermitian.
The two complex coefficients appear in two linearly independent combinations in the tensorial $Z$ boson and photon interactions. The real parts of these combinations, $\dvz$ and $\dva$, correspond to magnetic dipole moments, whereas the imaginary parts $\daz$, $\daa$ are CP-violating electric dipole moments. 

One can perform an additional simplification by noticing that the contribution from dimension-six operators to the $Zb_L b_L$ vertex~\cite{AguilarSaavedra:2008zc},
\begin{equation}
c_L^b = -1 - 2 s_W^2 Q_b + \left[  C_{\phi q}^{(3,3+3)} + C_{\phi q}^{(1,3+3)}
\right]  \frac{v^2}{\Lambda^2} \,,
\end{equation}
involves precisely the same operators as in the $Zt_L t_L$ vertex. (For $c_L^b$ the same normalisation as in Eq.~(\ref{ec:Ztt}) is used.) The bottom quark couplings have been probed with great precision at PETRA, LEP and SLD. Thus, given the precision that is expected for top couplings at the LHC and ILC, it is a good approximation to assume
\begin{equation}
C_{\phi q}^{(1,3+3)} \simeq -  C_{\phi q}^{(3,3+3)} \,,
\label{ec:Crel}
\end{equation}
since non-zero contributions from these operators must be balanced in order to keep the $Z b_L b_L$ vertex close to its SM value.
Besides, we point out that the exact equality between these coefficients automatically holds for some SM extensions, for example with new charge $2/3$ singlets~\cite{delAguila:1998tp,delAguila:2000aa}, so no fine-tuning is implied here.

We are now in position to compute the $t \bar t$ cross section at the ILC, including top trilinear operators.
The electron interactions have been probed with an excellent precision without noticing departures from the SM prediction. We then take them as in the SM,
\begin{equation}
\mathcal{L}_{e} = - e Q_e \bar e \, \gm  e\; A_\mu - \frac{g}{2 c_W} \bar e \, \gm \left( c_L^e P_L
  + c_R^e P_R \right) e\; Z_\mu \,, 
\end{equation}
with $c_L^e = -1 - 2 s_W^2 Q_e$, $c_R^e = -2  s_W^2 Q_e$ ($Q_e=-1$). 
In terms of these top quark and electron vertices, the polarised forward (F) and backward (B) cross sections for $e^+ e^- \to t \bar t$ are\footnote{We define as `forward' the events in which the top quark moves along the positron direction. The subindices in $e^+$, $e^-$ indicate the helicity. We keep quadratic terms in the operator coefficients, which is consistent with the $1/\Lambda^2$ expansion of the effective operator framework~\cite{AguilarSaavedra:2010sq}.}
\begin{small}
\begin{eqnarray}
\sigma_{F,B} (e_R^+ e_L^-) & = & \frac{\beta}{32 \pi} \left\{ s (3+\beta^2) \left[ |\vLL|^2 + |\vLR|^2 \right]  \mp 3 s \beta  \left[ |\vLL|^2 - |\vLR|^2 \right] + 24 m_t^2 \,\RE\, \vLL \vLR^*  \right. \notag \\
& & +2 s^2 (3-\beta^2) \left[ |\tLV|^2 + |\tLA|^2 \right]  +24 m_t^2 s \left[ |\tLV|^2 - |\tLA|^2 \right]  \notag \\
& & \left. -24s m_t \,  \RE \left[ (\vLL+\vLR) \tLV^* \right] \pm 12s m_t \beta \, \RE \left[ (\vLL-\vLR) \tLV^* \right]  \right\}  \,, \notag \\
\sigma_{F,B} (e_L^+ e_R^-) & = & \frac{\beta}{32 \pi} \left\{ s (3+\beta^2) \left[ |\vRL|^2 + |\vRR|^2 \right]  \pm 3 s \beta  \left[ |\vRL|^2 - |\vRR|^2 \right] + 24 m_t^2 \,\RE\, \vRL \vRR^*  \right. \notag \\
& & +2 s^2 (3-\beta^2) \left[ |\tRV|^2 + |\tRA|^2 \right]  +24 m_t^2 s \left[ |\tRV|^2 - |\tRA|^2 \right]  \notag \\
& & \left. -24s m_t \,  \RE \left[ (\vRL+\vRR) \tRV^* \right] \mp 12s m_t \beta \, \RE \left[ (\vRL-\vRR) \tRV^* \right]  \right\}  \,, \notag \\
\sigma_{F,B} (e_L^+ e_L^-) & = & \sigma_{F,B} (e_R^+ e_R^-) = 0 \,,
\end{eqnarray}
\end{small}%
with $\vij$, $\tij$ defined as
\begin{eqnarray}
\mathcal{V}_{ij} & = & e^2 \left[ \frac{c_i^e c_j^t}{4 s_W^2 c_W^2 (s-M_Z^2)} + \frac{Q_e Q_t}{s} \right] \,, \quad i,j=L,R \,, \notag \\
\mathcal{T}_{ij} & = & e^2 \left[ \frac{c_i^e d_j^Z}{4 s_W^2 c_W^2 M_Z (s-M_Z^2)} + \frac{Q_e d_j^\gamma }{s m_t} \right] \,, \quad i=L,R \,,~ j=V,A \,.
\end{eqnarray}
From these equations the cross sections and asymmetries for arbitrary electron (positron) polarisations $P_{e^-}$ ($P_{e^+}$)  can be straightforwardly obtained,
\begin{equation}
\sigma_{F,B} = \frac{1}{4} \left[ (1+P_{e^-}) (1-P_{e^+}) \sigma_{F,B}(e_L^+ e_R^-) +
(1-P_{e^-}) (1+P_{e^+}) \sigma_{F,B}(e_R^+ e_L^-) \right]
\end{equation}

Before our detailed analysis in section~\ref{sec:4}, it is clarifying to see from these equations how the different terms $\mathcal{V}_{ij}$, $\mathcal{T}_{ij}$ can be disentangled:
\begin{enumerate}
\item Both total cross sections and asymmetries have different dependence on  $\tLV$ and $\tRV$ (which interfere with the vector terms) and their axial counterparts (which do not).
\item Forward-backward (FB) asymmetries distinguish $\vLL$ from $\vLR$ and $\vRL$ from $\vRR$, and thus $c_L^t$ from $c_R^t$, for either beam polarisation, and hence also for unpolarised beams.
\item Beam polarisation distinguishes $\tLV$ from $\tRV$ and $\tLA$ from $\tRA$. (Also $\vLL$ from $\vRL$, and $\vLR$ from $\vRR$, but this is uninteresting for us because the differences between these arise from the left- and right-handed electron couplings, assumed here as in the SM.)  This allows to separate $d_j^Z$ from $d_j^\gamma$, because the former is multiplied by a parity-violating coupling and the latter by the electron charge.
\item Measurements at different CM energies can help resolve the vector ($\vij$) and tensor ($\tij$) contributions because the CM energy dependence is different. Note also that in the expressions of $\vij$ and $\tij$ the propagators are quite similar at ILC energies, $s-M_Z^2 \approx s$, so measurements at different CM energies cannot be used to distinguish off-shell $Z$ boson and photon contributions.
\end{enumerate}
 Initial state radiation and beamstrahlung modify  cross sections and asymmetries but, clearly, they do not affect the strategy to disentangle effective operator contributions. Our calculations do not take into account corrections from such effects, which fall beyond the scope of this work.

\section{ILC versus LHC sensitivity}
\label{sec:3}

\begin{figure}[htb]
\begin{center}
\begin{tabular}{ccc}
\epsfig{file=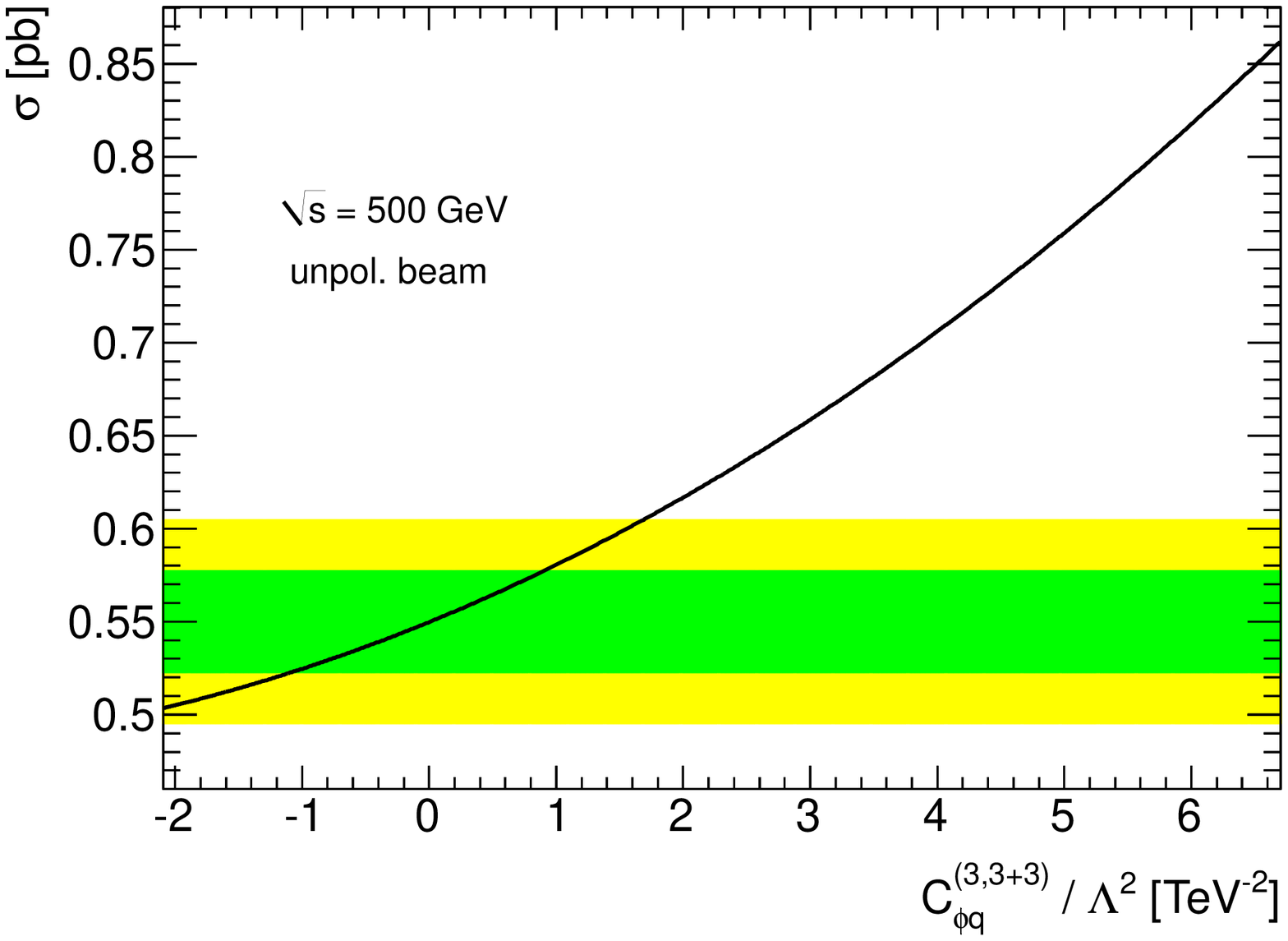,height=5.2cm,clip=} & \quad &
\epsfig{file=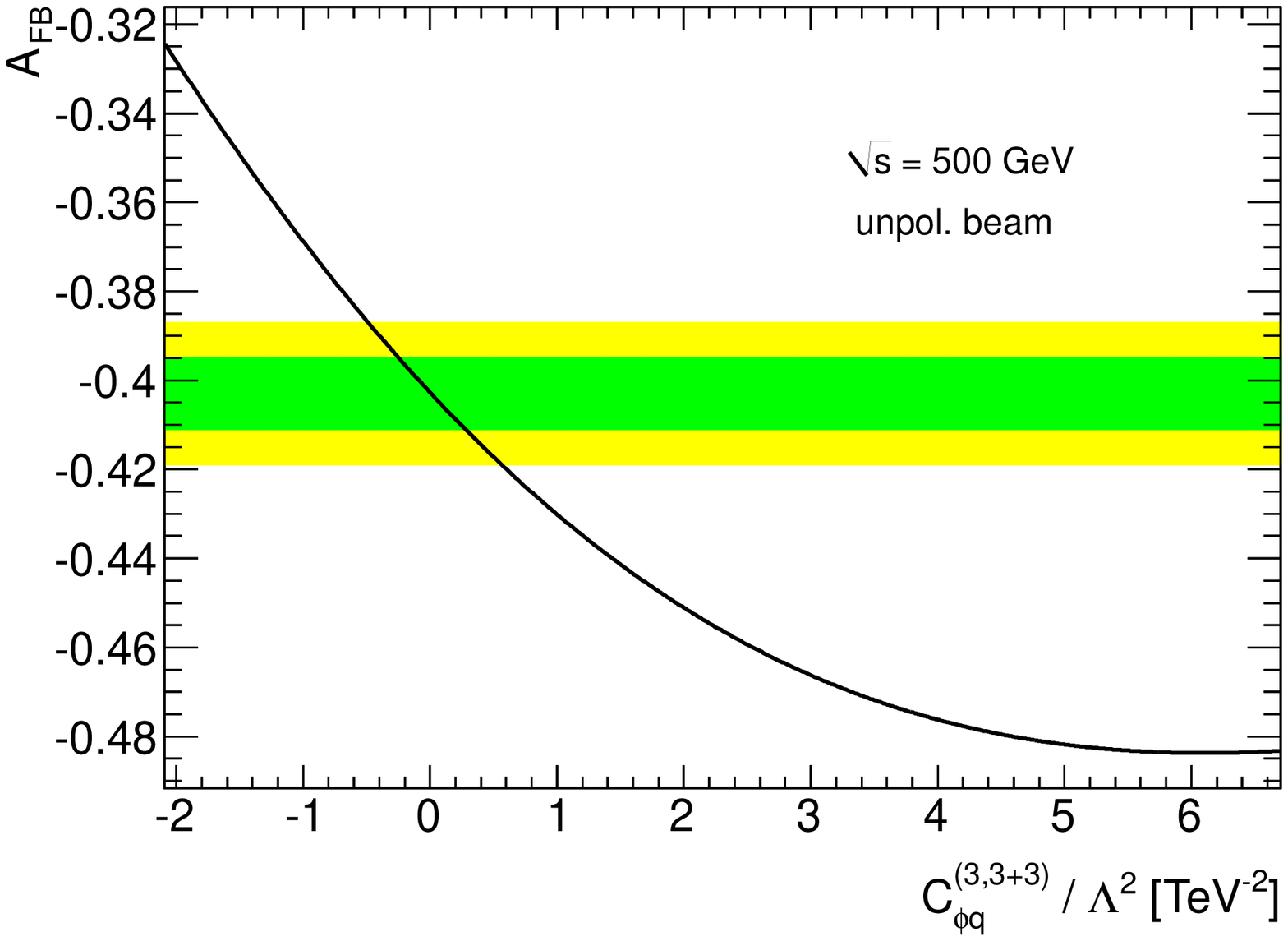,height=5.2cm,clip=}
\end{tabular}
\caption{Dependence of the unpolarised cross section and FB asymmetry on $C_{\phi q}^{(3,3+3)}$.}
\label{fig:sigma1}
\end{center}
\end{figure}

There are two effective operators involved in the $Ztt$ vertex which have already been probed at the LHC: 
$O_{\phi q}^{(3,3+3)}$ and $O_{uW}^{33}$. Both operators also modify the $Wtb$ vertex, which can be parameterised as~\cite{AguilarSaavedra:2008zc}
\begin{equation}
\mathcal{L}_{Wtb} = - \frac{g}{\sqrt 2} \bar b \, \gamma^{\mu} \left( \vl P_L + \vr P_R
\right) t\; W_\mu^- - \frac{g}{\sqrt 2} \bar b \, \frac{i \sigma^{\mu \nu} q_\nu}{M_W}
\left( \gl P_L + \gr P_R \right) t\; W_\mu^- + \mathrm{h.c.} \,,
\label{ec:lagr}
\end{equation}
with
\begin{align}
& \vl = V_{tb} + C_{\phi q}^{(3,3+3)} \frac{v^2}{\Lambda^2} \,, 
&& \gl = \sqrt 2 C_{dW}^{33*} \frac{v^2}{\Lambda^2} \,, \notag \\
& \vr = \frac{1}{2} C_{\phi \phi}^{33*} \frac{v^2}{\Lambda^2} \,,
&& \gr = \sqrt 2 C_{uW}^{33} \frac{v^2}{\Lambda^2} \,.
\end{align}
$O_{\phi \phi}^{33}$ and $O_{dW}^{33}$ are two other top trilinear operators which do not mediate neutral interactions. Limits on $C_{\phi q}^{(3,3+3)}$ can be extracted from single top cross section measurements. For example, from the ATLAS $t$-channel measurement~\cite{Aad:2012ux} one can get the limit
\begin{equation}
\frac{C_{\phi q}^{(3,3+3)}}{\Lambda^2} \in [-2.1,6.7]~\text{TeV}^{-2}
\end{equation}
with a 95\% confidence level (CL), assuming no other non-SM contribution to single top production. The variation of the unpolarised cross section and FB asymmetry at ILC for $C_{\phi q}^{(3,3+3)}$ ranging in this interval is presented in Fig.~\ref{fig:sigma1}. The CM energy is taken as $\sqrt s = 500$ GeV. The bands represent a $1\sigma$ (inner, green) and $2\sigma$ (outer, yellow) variation around the SM value, assuming total uncertainties of 5\% in the cross section and 2\% in the asymmetry~\cite{Doublet:2012wf}.\footnote{To our knowledge there are not yet complete studies of experimental systematics in the $t \bar t$ cross section and asymmetry measurements, and these values seem a reasonable estimate, given the expected improvement over LHC systematics for an $e^+ e^-$ machine.  (Statistical uncertainties for cross sections and asymmetries are below 1\% already for a luminosity of 100 fb$^{-1}$.) The main results of this paper, that is, the improvement with respect to the LHC sensitivity and the possibility of disentangling effective operator contributions, are largely independent of the precise numbers assumed. }  Here the rest of operator coefficients are assumed zero, except for the relation in Eq.~(\ref{ec:Crel}). The improvement of the ILC with respect to the LHC is evident, and comes not only from the smaller cross section uncertainties at the ILC but also because the contribution of this operator is enhanced via Eq.~(\ref{ec:Crel}).

\begin{figure}[htb]
\begin{center}
\begin{tabular}{ccc}
\epsfig{file=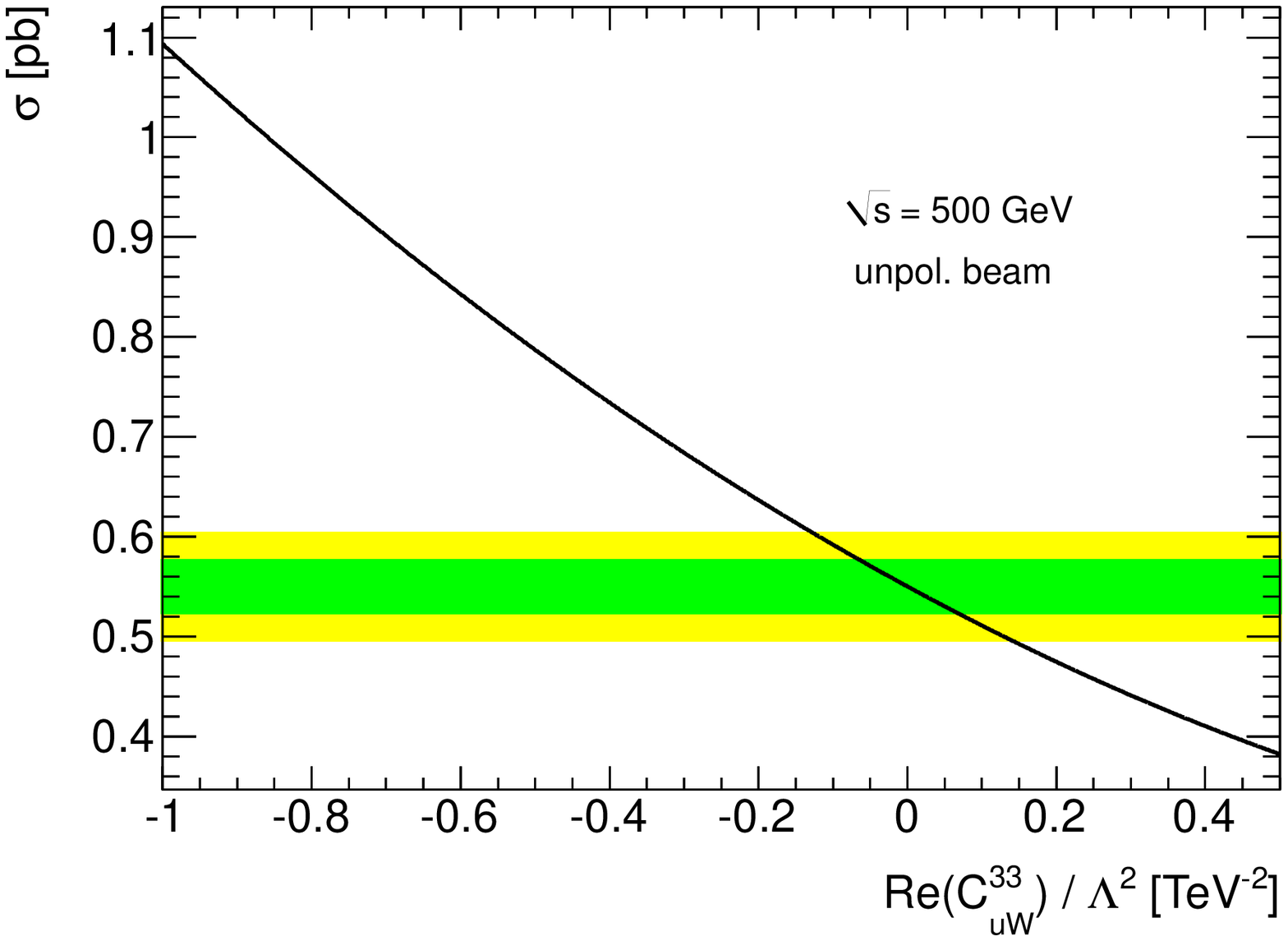,height=5.2cm,clip=} & \quad &
\epsfig{file=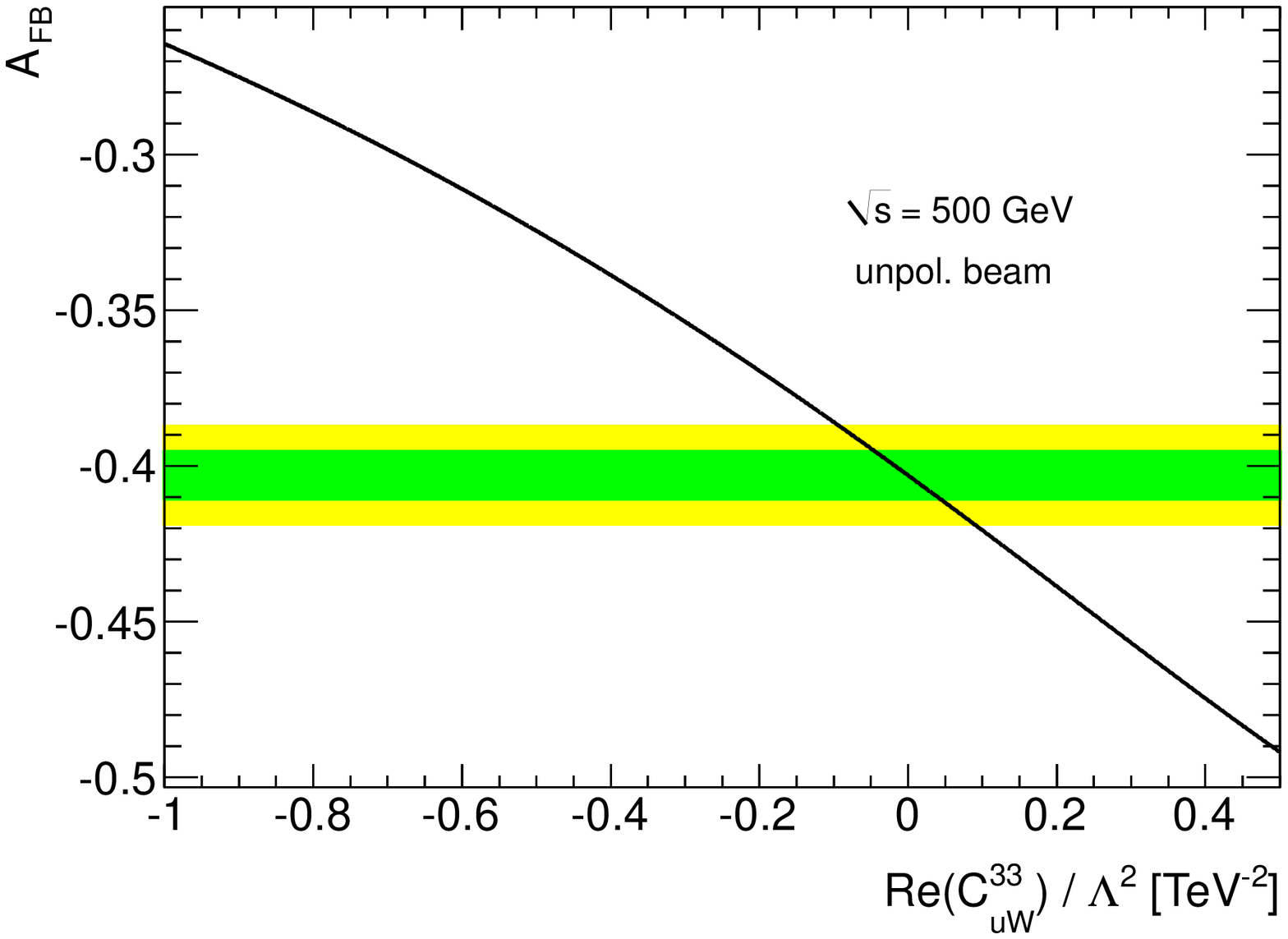,height=5.2cm,clip=}
\end{tabular}
\caption{Dependence of the unpolarised cross section and FB asymmetry on $\RE\,C_{uW}^{33}$.}
\label{fig:sigma2}
\end{center}
\end{figure}

Limits on $C_{uW}^{33}$ have already been obtained from the measurement of helicity fractions in top decays by the ATLAS Collaboration~\cite{Aad:2012ky},
\begin{equation}
\frac{\RE \, C_{uW}^{33}}{\Lambda^2} \in [-1.0,0.5]~\text{TeV}^{-2} \,,
\end{equation}
also with a 95\% CL. In Fig.~\ref{fig:sigma2} we plot the variation of the unpolarised cross section and FB asymmetry at ILC, for $\RE \, C_{uW}^{33}$ within these limits. For this operator, the excellent sensitivity mainly stems from the $\sqrt s/m_t$ enhancement of its contribution to $e^+ e^- \to t \bar t$ with respect to $W$ helicity observables~\cite{AguilarSaavedra:2006fy}. Assuming that the operator coefficient equals unity, the sensitivity to the new physics scale $\Lambda$ extends up to 4.5 TeV for this CM energy.

\begin{figure}[htb]
\begin{center}
\begin{tabular}{ccc}
\epsfig{file=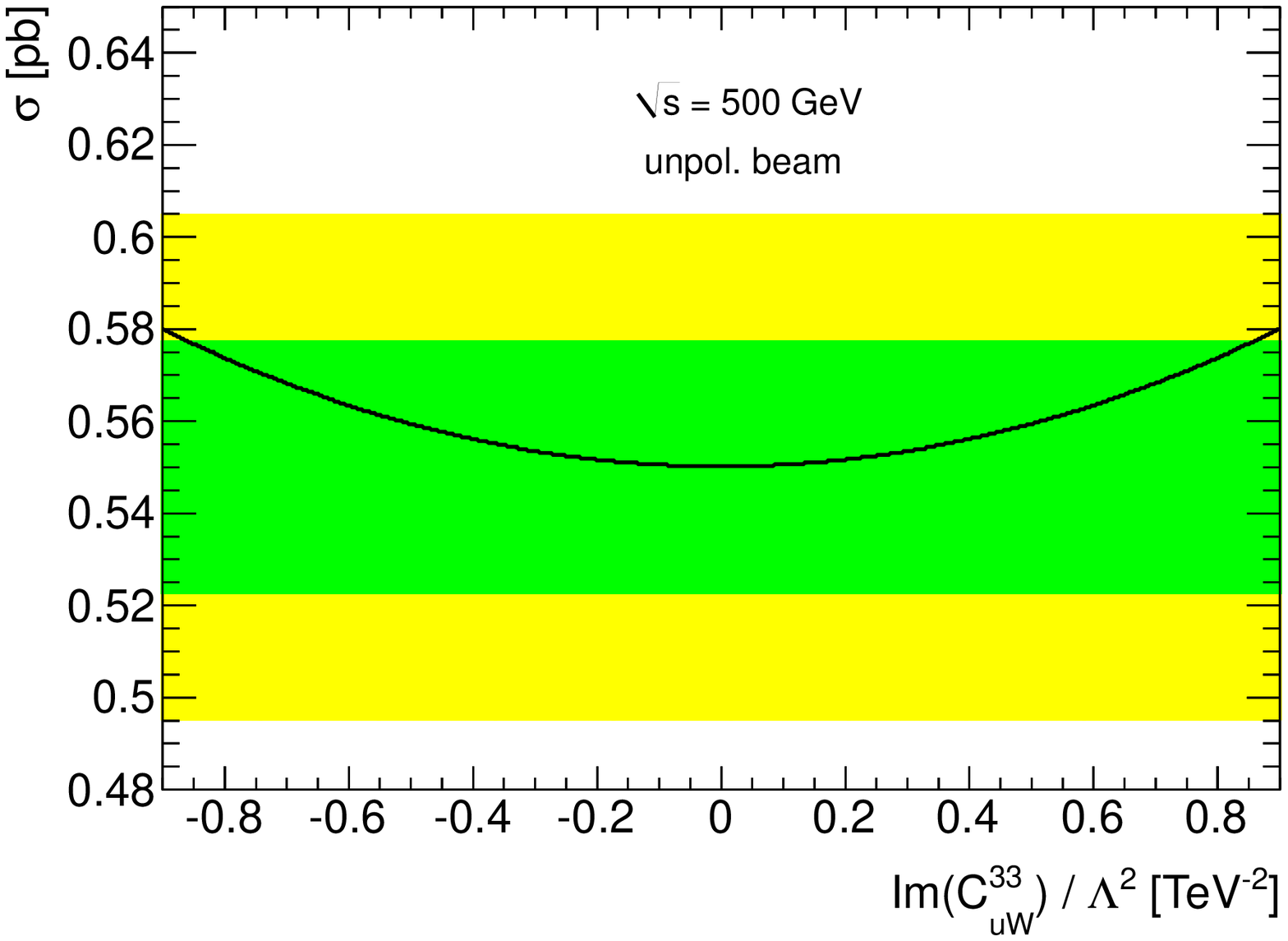,height=5.2cm,clip=} & \quad &
\epsfig{file=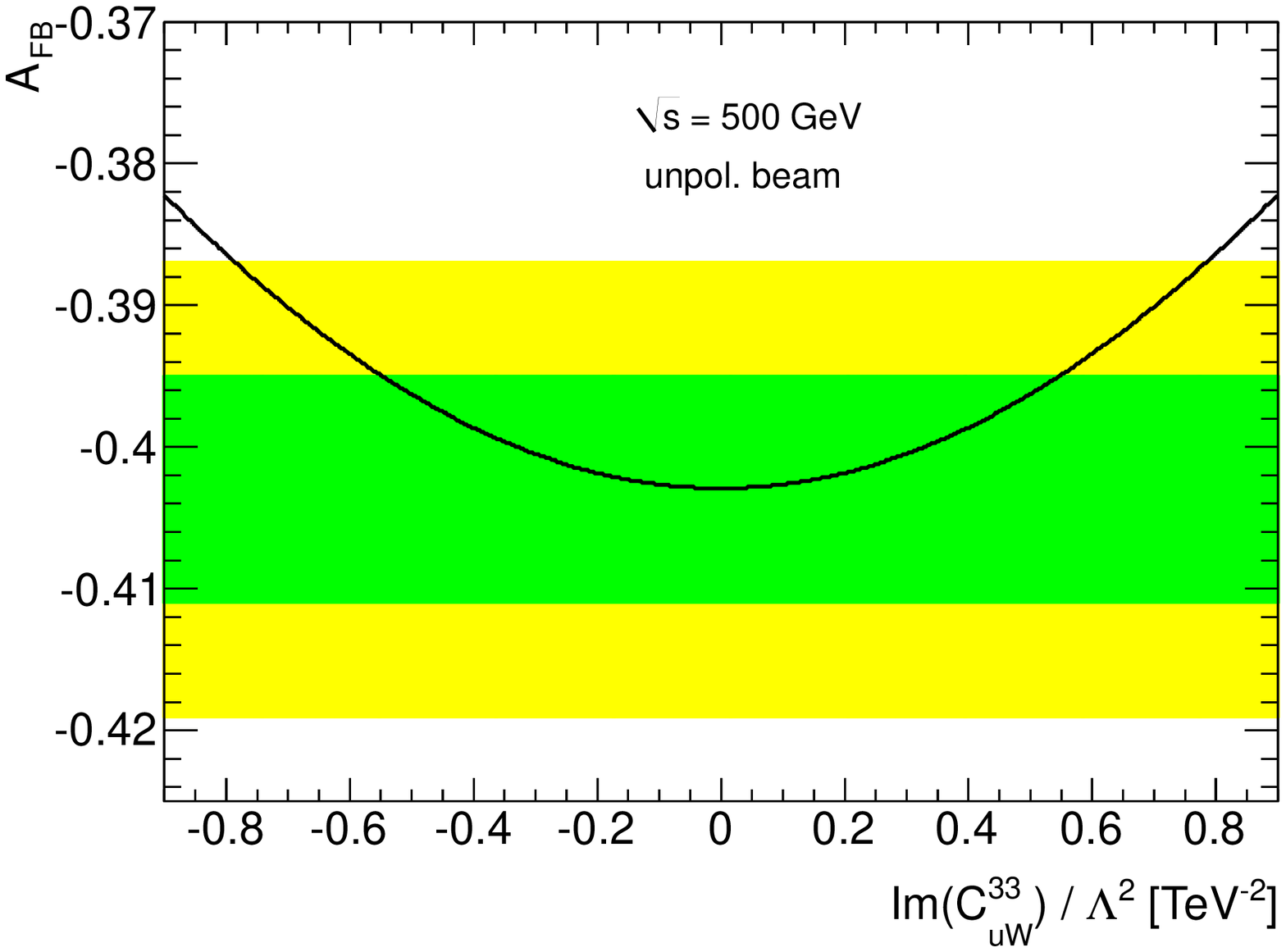,height=5.2cm,clip=}
\end{tabular}
\caption{Dependence of the unpolarised cross section and FB asymmetry on $\IM\,C_{uW}^{33}$.}
\label{fig:sigma3}
\end{center}
\end{figure}

The anti-Hermitian part of this operator can also be probed with a CP-violating asymmetry $A_\text{FB}^N$ defined for polarised top decays~\cite{AguilarSaavedra:2010nx}, being the estimated sensitivity
\begin{equation}
\frac{\IM \, C_{uW}^{33}}{\Lambda^2} \in [-0.9,0.9]~\text{TeV}^{-2} \,.
\end{equation}
The corresponding variation of the $t \bar t$ cross section and asymmetry are shown in Fig.~\ref{fig:sigma3}. The sensitivity is moderate in this case and comparable to the one at the LHC, in spite of the fact that the anti-Hermitian part of this operator does not interfere with the SM in CP-conserving quantities such as total cross sections and asymmetries, and their dependence on $\IM\, C_{uW}^{33}$ is quadratic (as it can be readily observed in the plots).

\begin{figure}[htb]
\begin{center}
\begin{tabular}{ccc}
\epsfig{file=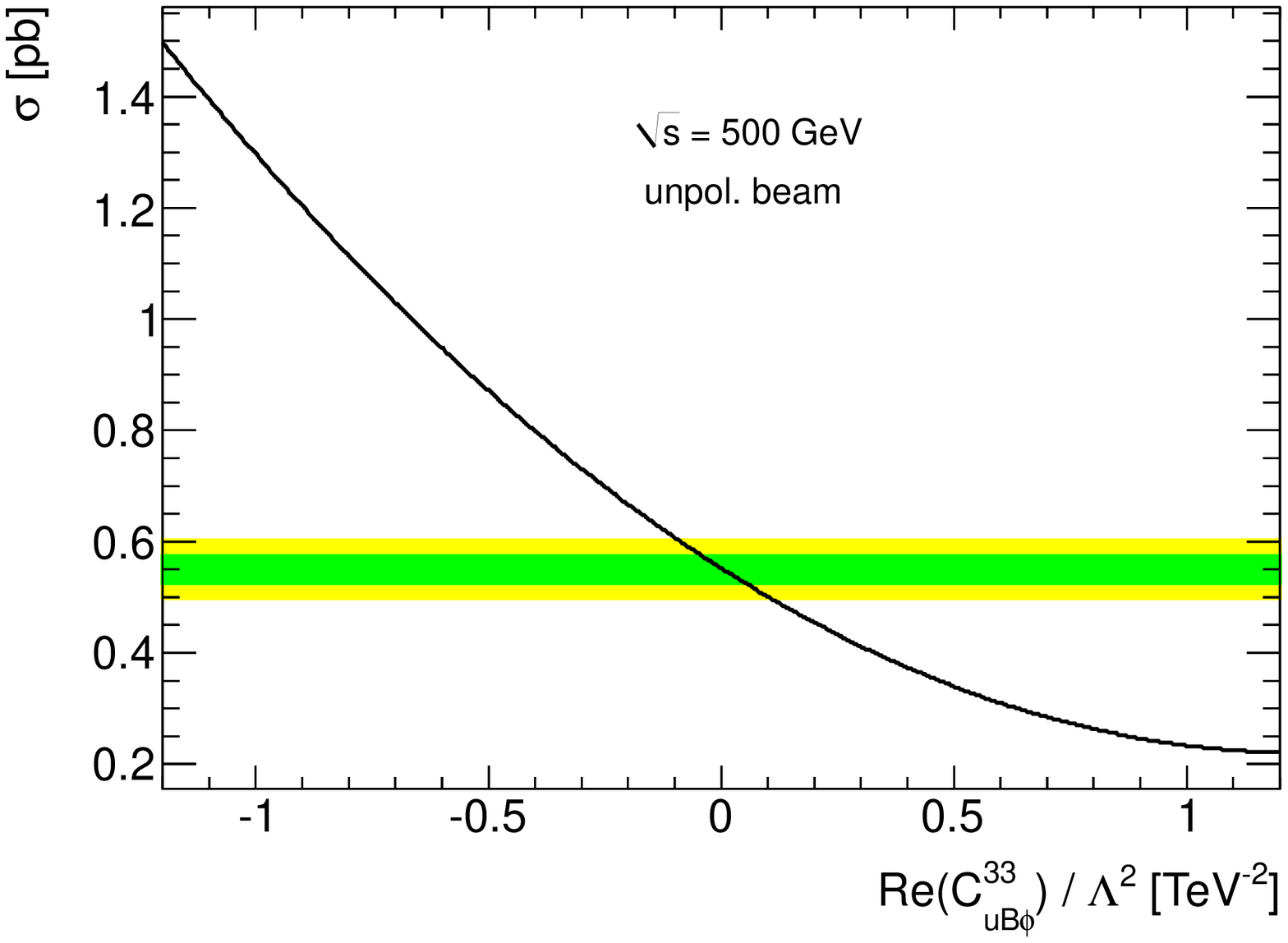,height=5.2cm,clip=} & \quad &
\epsfig{file=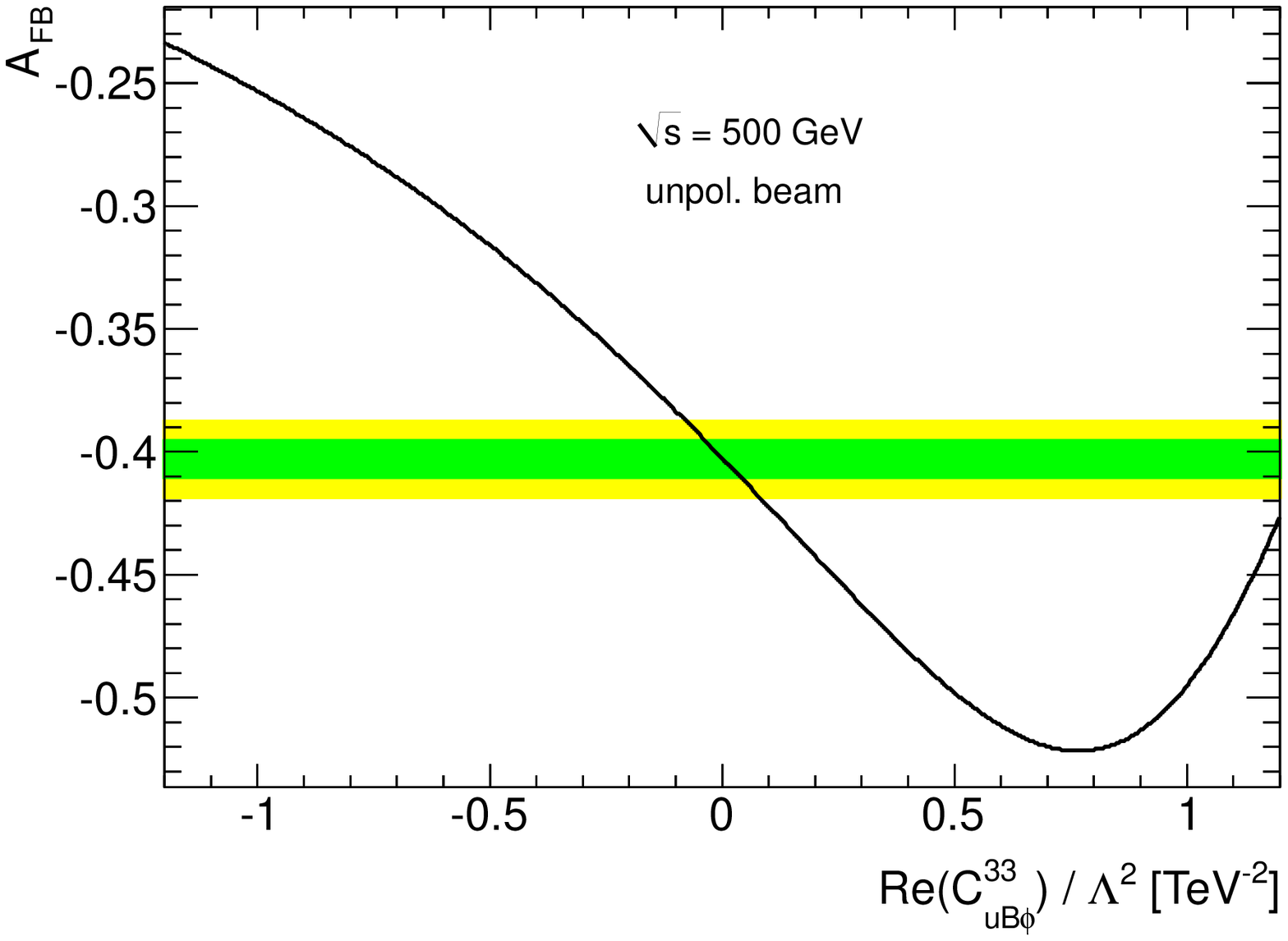,height=5.2cm,clip=} \\[1mm]
\epsfig{file=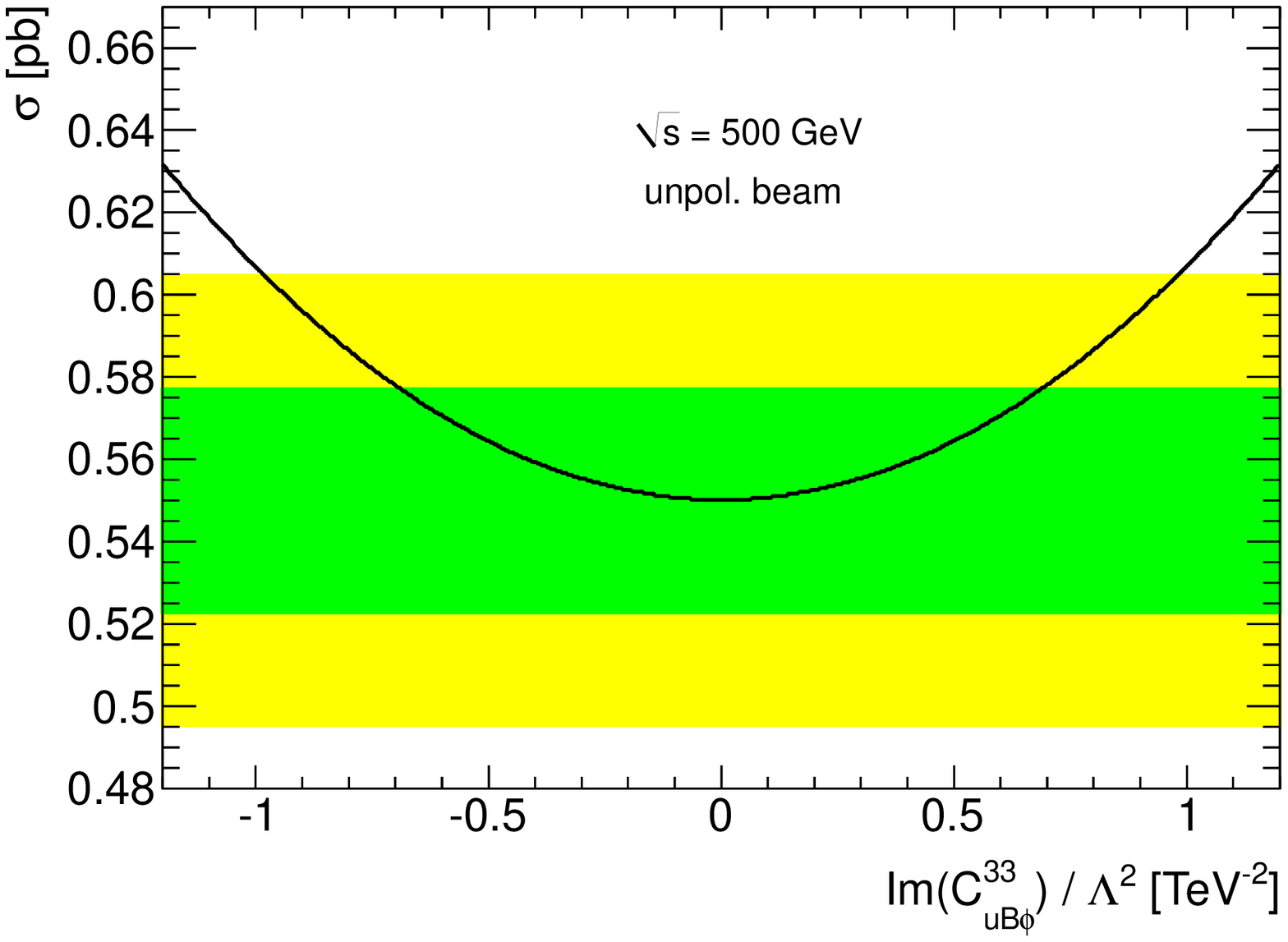,height=5.2cm,clip=} & \quad &
\epsfig{file=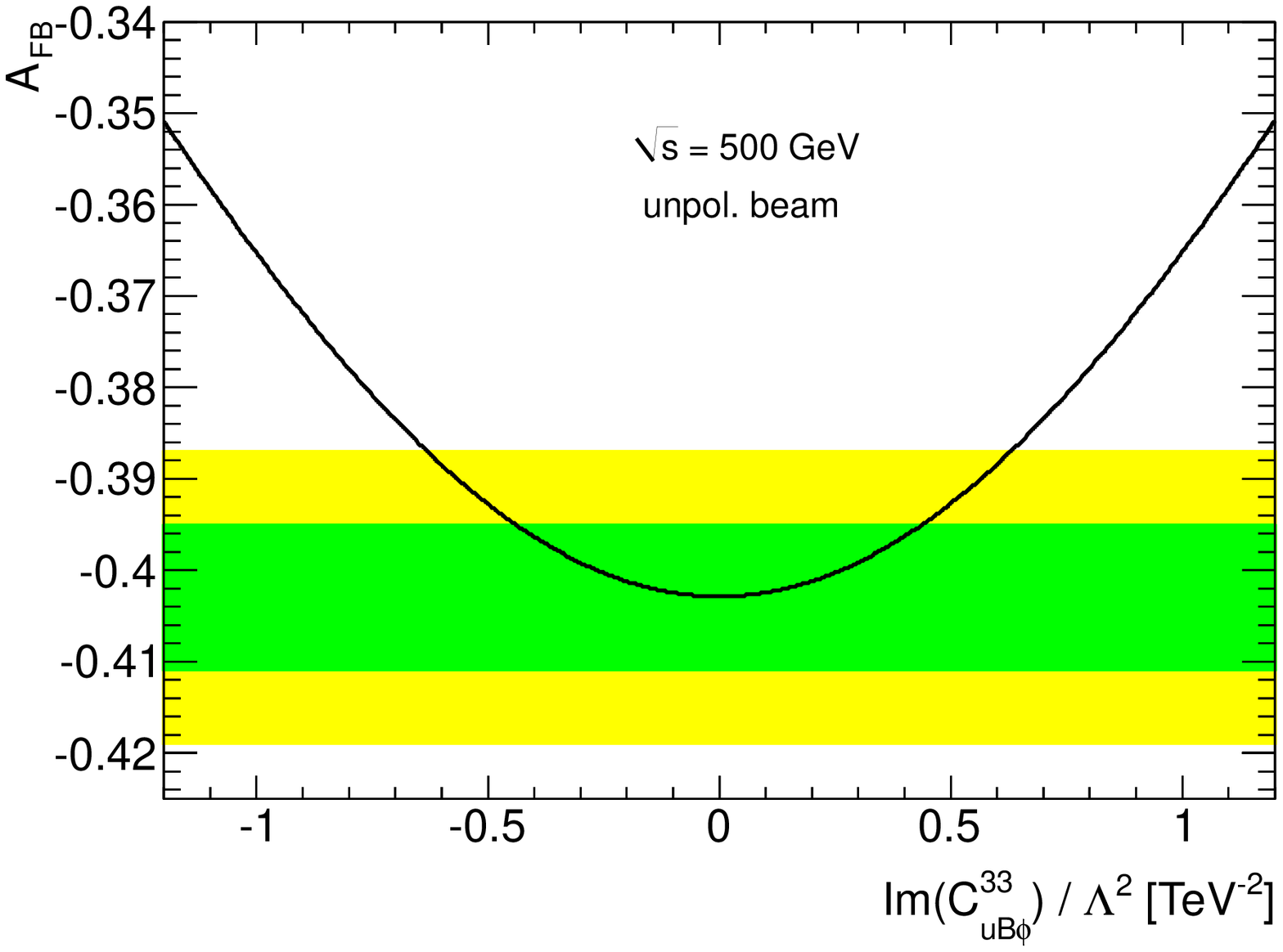,height=5.2cm,clip=} 
\end{tabular}
\caption{Dependence of the unpolarised cross section and FB asymmetry on $\RE\,C_{uB\phi}^{33}$ (up) and $\IM\,C_{uB\phi}^{33}$ (down).}
\label{fig:sigma4}
\end{center}
\end{figure}

Future LHC limits have also been estimated for $t \bar t Z$ and $t \bar t \gamma$ production~\cite{Baur:2004uw}. The best ones are for the latter process, and translated into our framework give
\begin{equation}
\frac{\RE\,C_{uW}^{33}}{\Lambda^2} \,, \frac{\IM\,C_{uW}^{33}}{\Lambda^2}  \in [-2.1,2.1]~\text{TeV}^{-2} \,, \quad \frac{\RE \, C_{uB\phi}^{33}}{\Lambda^2} \,, \frac{\IM \,C_{uB\phi}^{33}}{\Lambda^2} 
\in [-1.2,1.2]~\text{TeV}^{-2} \,.
\end{equation}
The potential limit on $\RE \, C_{uW}^{33}$ has already been surpassed by the ATLAS $W$ helicity measurement~\cite{Aad:2012ky} and the limit on $\IM \, C_{uW}^{33}$ from CP violation in top decays is expected to be better. On the other hand, potential LHC limits on $C_{uB\phi}^{33}$ are relevant but would be surpassed at the ILC, as it can be seen in Fig.~\ref{fig:sigma4}.

\section{Disentangling operator contributions}
\label{sec:4}

At the LHC, the $Ztt$ and $\gamma tt$ couplings can be independently measured in $t \bar tZ$ and $t \bar t\gamma$ associated production, respectively. However, the ILC sensitivity to anomalous contributions is much better, posing the question of whether the different contributions can also be disentangled at this collider, given the fact that both $Z,\gamma$ exchange in the $s$ channel contribute to $e^+ e^- \to t \bar t$.\footnote{Properly speaking, for tensor couplings the issue is not to measure separately the photon and $Z$ boson couplings, but to disentangle possible contributions from the two operators $O_{uW}^{33}$ and $O_{uB\phi}^{33}$, which simultaneously contribute to the $Ztt$ and $\gamma tt$ vertices with different weights, see Eqs.~(\ref{ec:gatt2}). In this sense, the usual assumption of setting either the photon or $Z$ boson contribution to zero to obtain limits on the other, is not useful in our effective operator framework.} This is possible by using the different options proposed for the ILC, like beam polarisation and a CM energy upgrade to 1 TeV. We assume that an electron longitudinal polarisation $P_e = \pm 0.8$ is possible. Additional positron polarisation improves our results, but, since this possibility is still under debate, we do not make use of it. We will not use the left-right asymmetry $A_{LR}$ as a constraint, because it is not independent from the polarised cross sections already considered. However, we note that experimental systematics may be smaller for this observable, and in practice it may be useful to include it too. The limits we present here do not result from a global fit but they are obtained requiring $1\sigma$ agreement of the different cross sections and FB asymmetries considered in each case.

\begin{figure}[htb]
\begin{center}
\begin{tabular}{ccc}
\epsfig{file=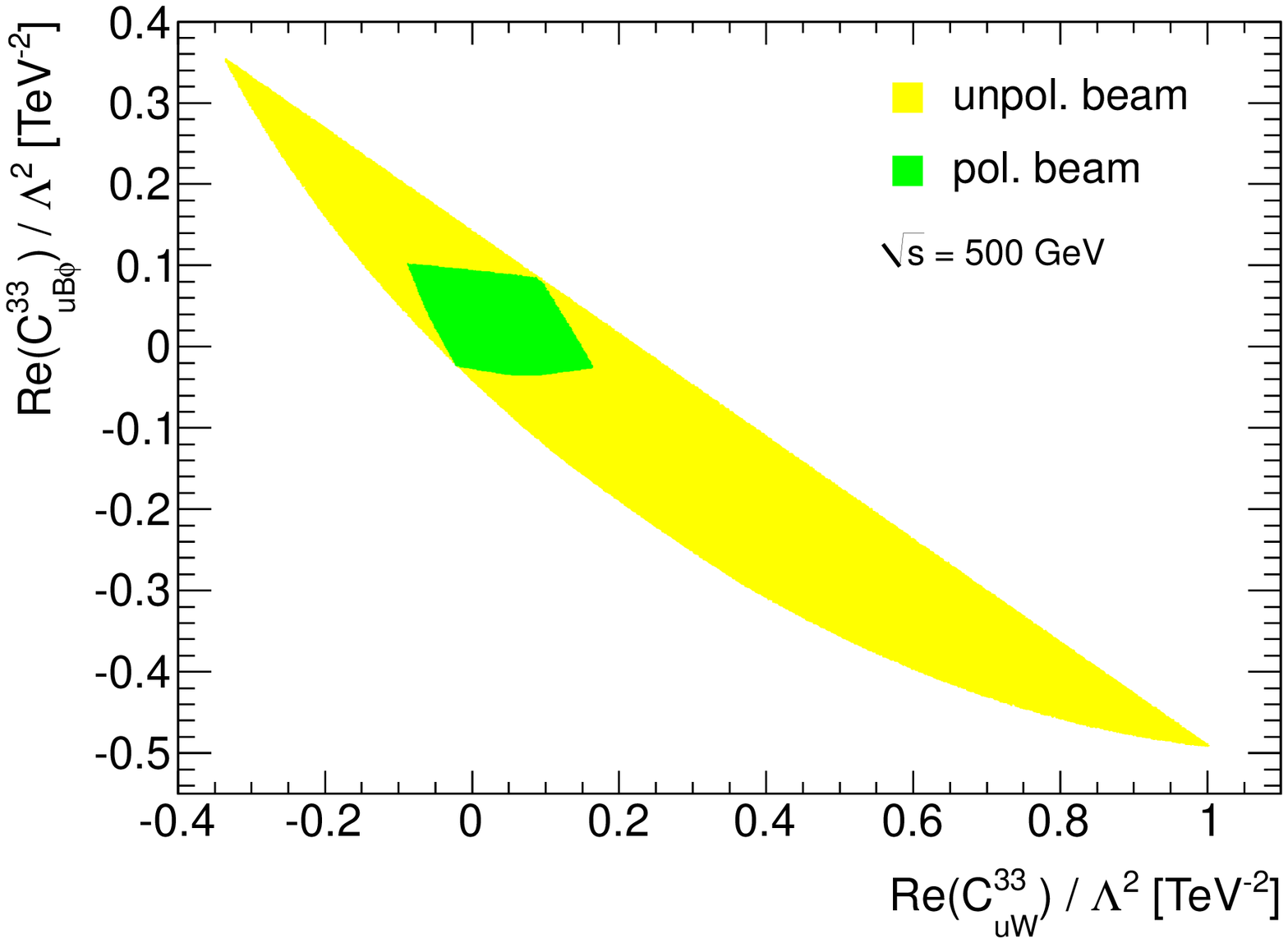,height=5.2cm,clip=} & \quad &
\epsfig{file=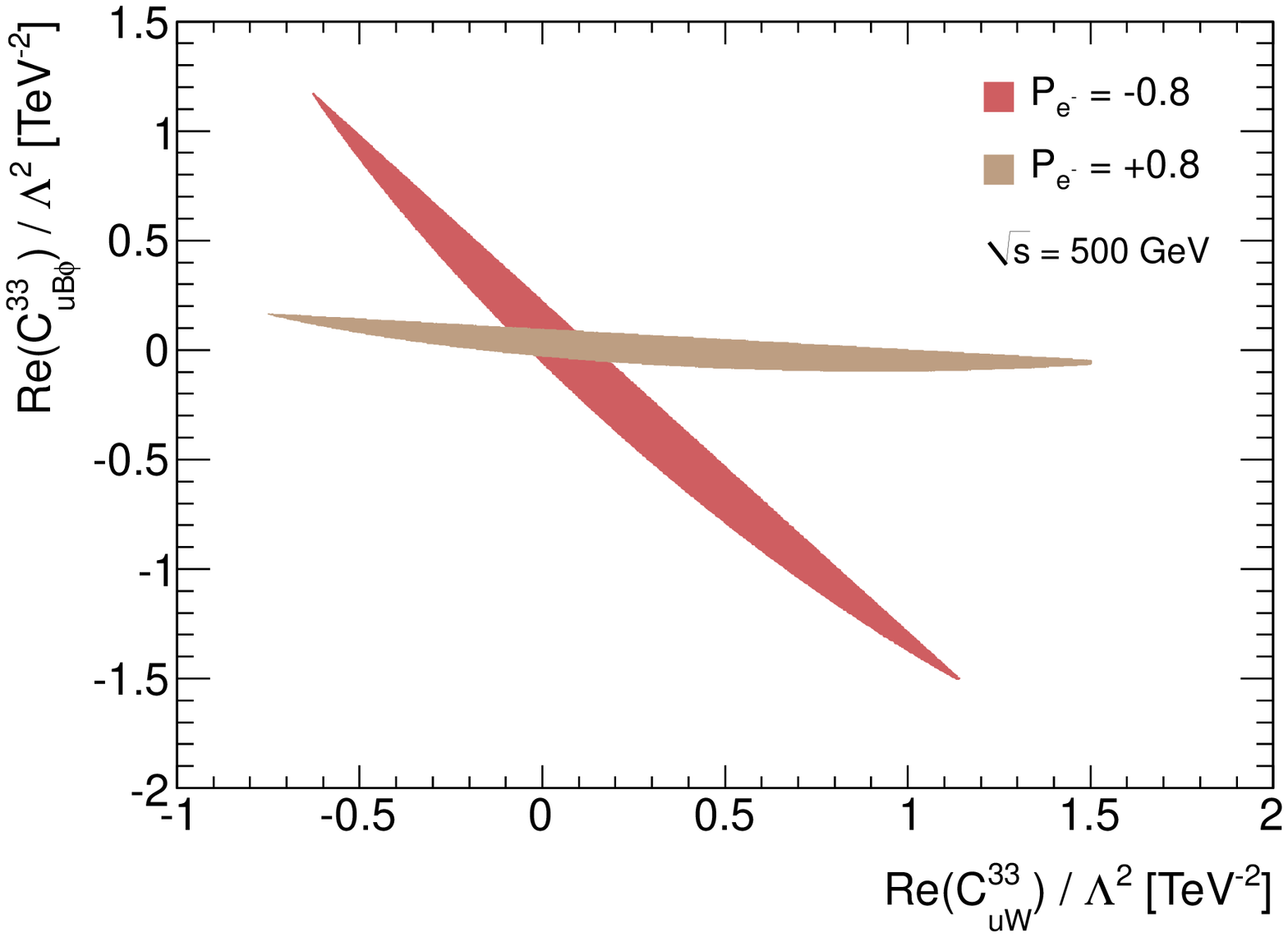,height=5.2cm,clip=}
\end{tabular}
\caption{Left: combined limits on $C_{uW}^{33}$ and $C_{uB\phi}^{33}$ for the cases of no beam polarisation and electron beam polarisation (only the real parts of these coefficients are shown). Right: complementarity of the measurements for $P_{e^-} = 0.8$ and $P_{e^-} = - 0.8$.}
\label{fig:pol}
\end{center}
\end{figure}

Measurements performed with different electron polarisations allow to distinguish $d_i^Z$ from $d_i^\gamma$, $i=V,A$, which in turn allows to disentangle $C_{uW}^{33}$ and $C_{uB\phi}^{33}$. To illustrate this, we use a simplified scenario where the rest of operator coefficients are set to zero, while these two are assumed complex. We show in Fig.~\ref{fig:pol} (left) the combined limits on $\RE \, C_{uW}^{33}$ and $\RE \, C_{uB\phi}^{33}$ without and with beam polarisations. In the unpolarised case (yellow region) the  measurements of both coefficients are largely anti-correlated, while the use of electron polarisation (green region) allows to determine the both quantities with a far smaller uncertainty. This great improvement results from the complementarity of limits for left- and right-handed beams, whose corresponding allowed regions are nearly orthogonal (right panel).

\begin{figure}[htb]
\begin{center}
\epsfig{file=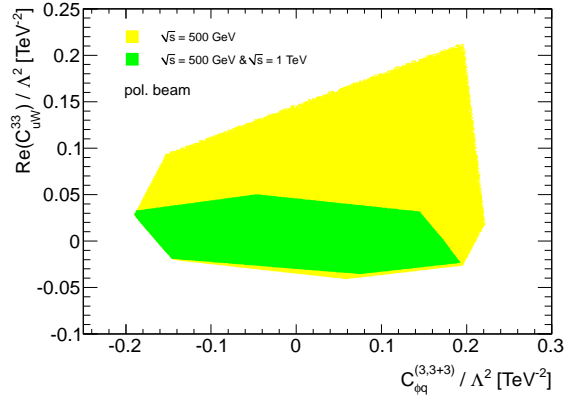,height=5.2cm,clip=}
\caption{Combined limits on $C_{\phi q}^{(3,3+3)}$ and $C_{uW}^{33}$ for a CM energy of 500 GeV and also with 1 TeV (only the real part of the latter coefficient is shown).}
\label{fig:E1000}
\end{center}
\end{figure}

As previously pointed out, measurements taken at different CM energies allow to distinguish $\gamma^\mu$ and $\sigma^{\mu \nu}$ couplings. At a given CM energy, it is often possible to fine-tune a cancellation between their contributions to cross sections and asymmetries so that the overall effects are small. This is not possible, however, at different CM energies, such as 500 GeV and 1 TeV, because the energy dependence of these contributions is different. An example of this interplay is shown in Fig.~\ref{fig:E1000}, where we consider a simplified scenario where only $C_{\phi q}^{(3,3+3)}$ and $C_{uW}^{33}$ are non-zero. The yellow region corresponds to limits with polarised beams at 500 GeV only, whereas the green region also includes limits at 1 TeV. 

\begin{figure}[htb]
\begin{center}
\begin{tabular}{ccc}
\epsfig{file=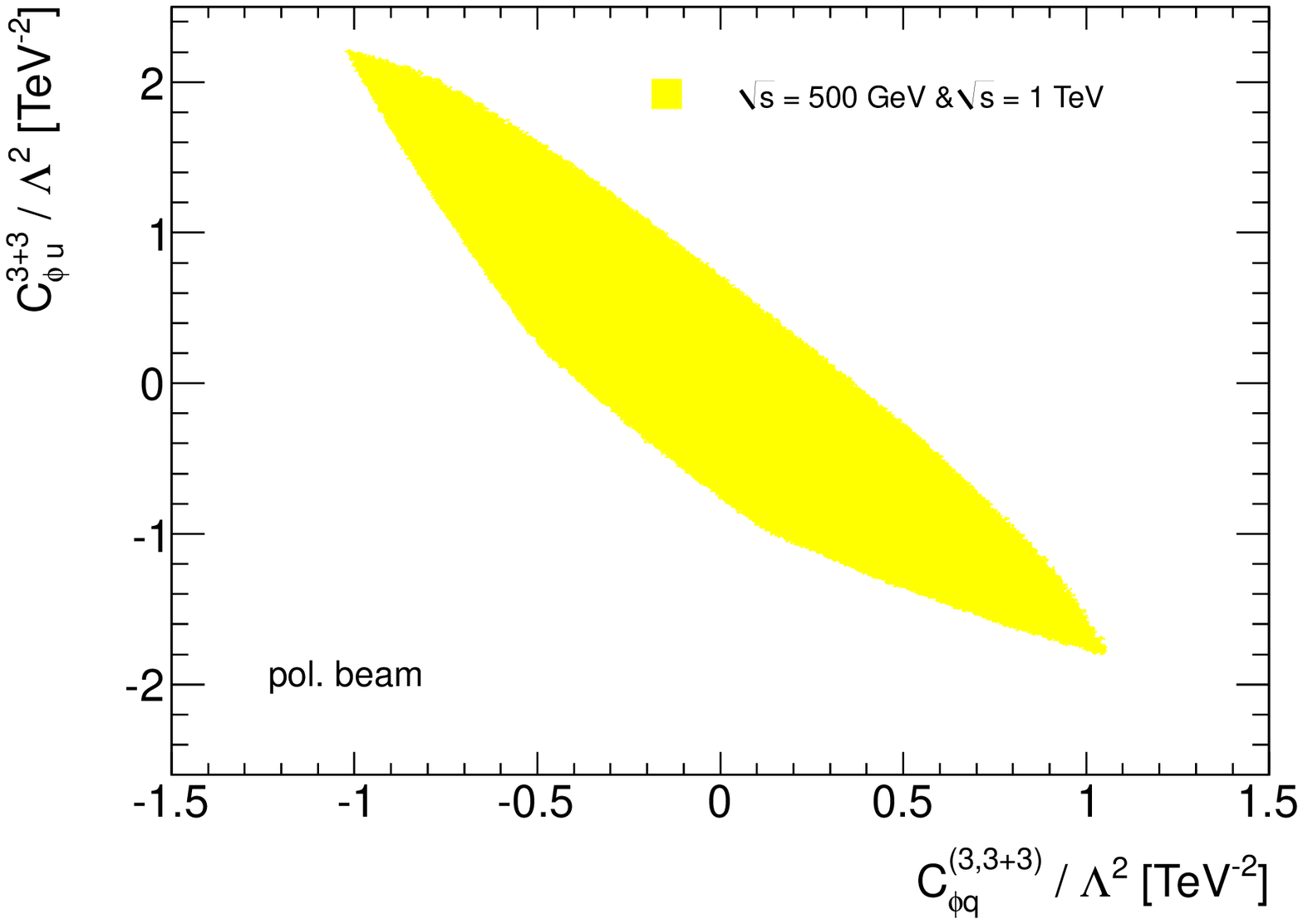,height=5.2cm,clip=} & \quad & 
\epsfig{file=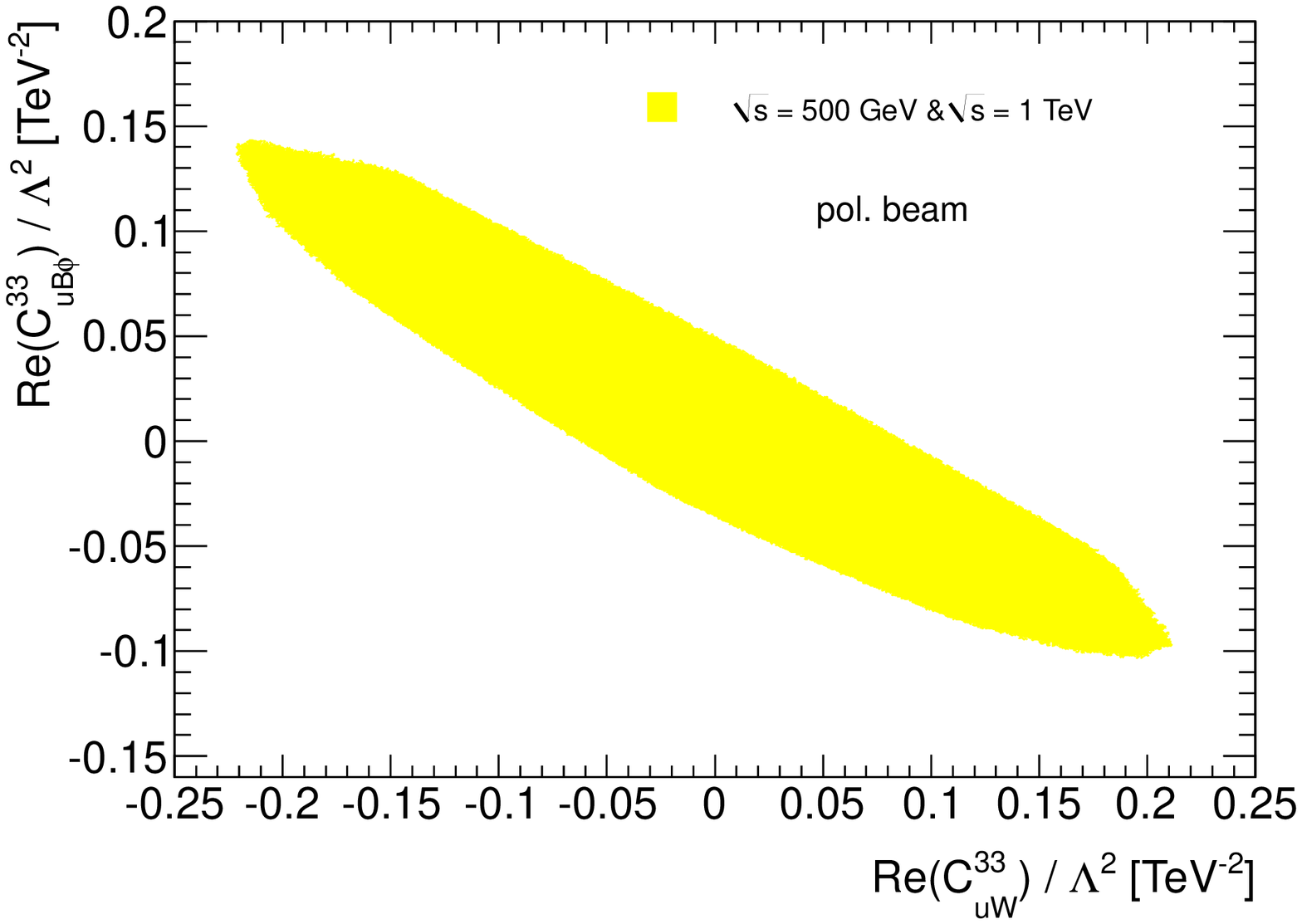,height=5.2cm,clip=} \\[1mm]
\multicolumn{3}{c}{\epsfig{file=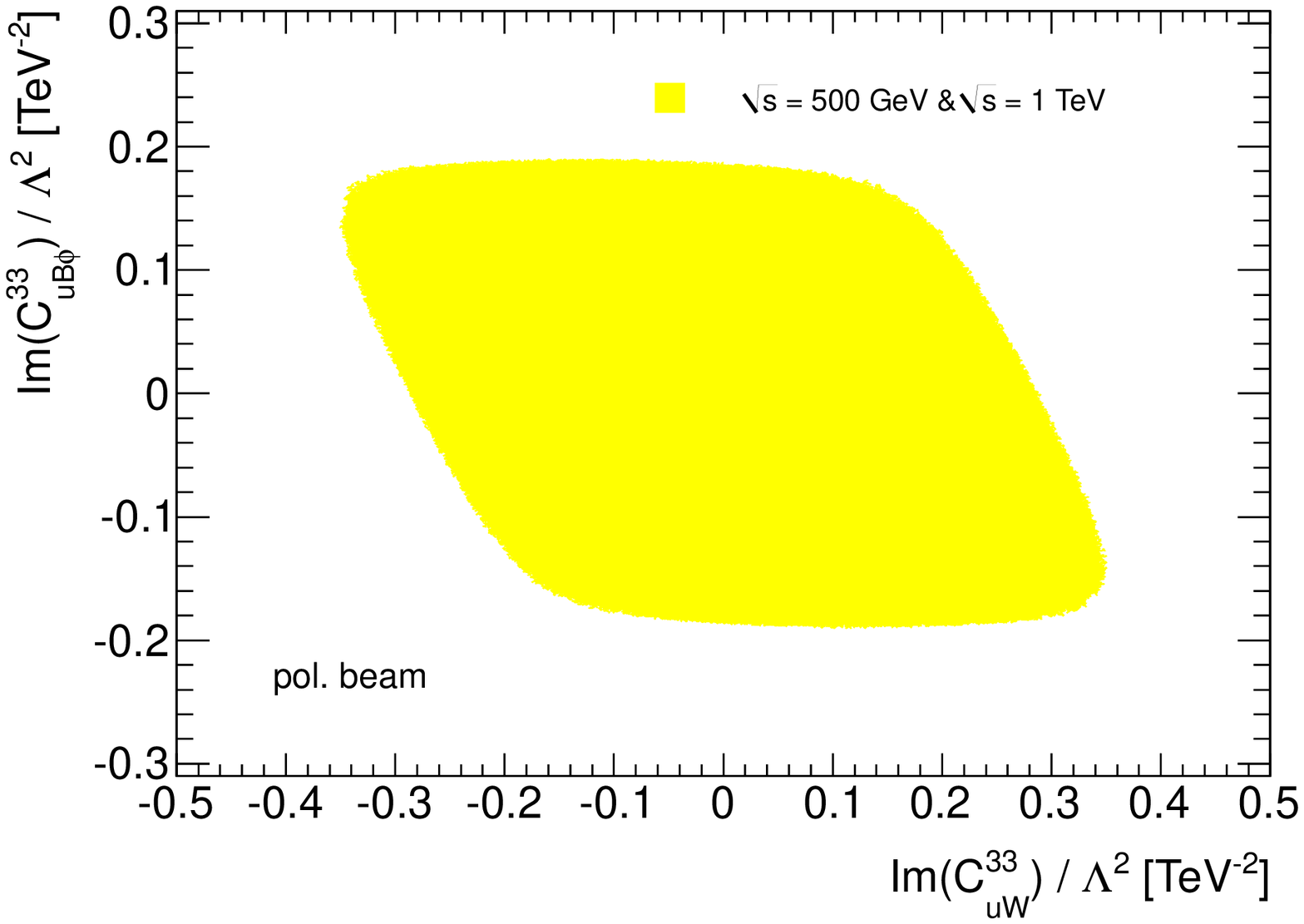,height=5.2cm,clip=}}
\end{tabular}
\caption{Combined limits on $Ztt$ and $\gamma tt$ trilinear effective operator coefficients.}
\label{fig:comb}
\end{center}
\end{figure}

Having shown the complementarity of the different beam polarisation and CM energy options, we show in Fig~\ref{fig:comb} the general limits for arbitrary $C_{\phi q}^{(3,3+3)}$, $C_{\phi u}^{3+3}$, $C_{uW}^{33}$  and $C_{uB\phi}^{33}$, the latter two complex (six real parameters in total), using polarised cross section and asymmetry measurements at 500 GeV and 1 TeV (eight constraints in total). These limits are excellent for $C_{uW}^{33}$ and $C_{uB\phi}^{33}$, even if there is a large anti-correlation between the limits on their real parts. For  $C_{\phi q}^{(3,3+3)}$ and $C_{\phi u}^{3+3}$ the limits are also interesting and better than the ones expected at the LHC through measurements of the single top cross section. We remark here that these combined limits are numerically worse than the sensitivities shown in section~\ref{sec:3} because here we allow for all possible cancellations between operator contributions. For example, if we set $C_{\phi u}^{3+3}$ to zero, the resulting limit on $C_{\phi q}^{(3,3+3)}$ improves by more than a factor of two.

\section{Summary}
\label{sec:5}

In this paper we have investigated the effect of top trilinear effective operators in $e^+ e^- \to t \bar t$ at ILC energies. Our first purpose has been to investigate ILC the sensitivity to these operators, comparing with the LHC. It is already known that the sensitivity to $Ztt$ and $\gamma tt$ couplings is better at the ILC than in $t \bar t Z$ and $t \bar t \gamma$ at the LHC~\cite{Baur:2004uw}. But, at variance with previous approaches, the effective operator framework adopted also allows for a direct comparison with charged current processes at the LHC, like single top production and decays $t \to Wb$. We have found that, despite the fact that the LHC prospects are already good due to its excellent statistics, the ILC sensitivity is even better for those operators. Assuming operator coefficients equal to unity, the new physics scales probed extend up to 4.5 TeV, for a CM energy of 500 GeV.

A second issue we have investigated in detail is how to set simultaneous bounds on all the operators involved, which contribute to the $Ztt$ and $\gamma tt$ vertices. We have shown that the use of electron beam polarisation is essential to disentangle contributions, as is the combination of measurements at 500 GeV and 1 TeV. The results presented here make manifest that the determination of top interactions constitute a physics case for the use of electron beam polarisation, as well as for a possible CM energy upgrade to 1 TeV.

\section*{Acknowledgements}
This work has been supported by MICINN by projects FPA2006-05294 and FPA2010-17915, Junta de Andaluc\'{\i}a (FQM 101, FQM 03048 and FQM 6552) and Funda\c c\~ao
para a Ci\^encia e Tecnologia~(FCT) project CERN/FP/123619/2011. 
The work of M.C.N.~Fiolhais has been supported by FCT grant SFRH/BD/48680/2008.

\end{document}